\documentclass{article}

\usepackage{PRIMEarxiv}

\usepackage[utf8]{inputenc} % allow utf-8 input
\usepackage[T1]{fontenc}    % use 8-bit T1 fonts
\usepackage{hyperref}       % hyperlinks
\usepackage{url}            % simple URL typesetting
\usepackage{booktabs}       % professional-quality tables
\usepackage{amsfonts}       % blackboard math symbols
\usepackage{nicefrac}       % compact symbols for 1/2, etc.
\usepackage{microtype}      % microtypography
\usepackage{lipsum}
\usepackage{graphicx}
\usepackage{amsmath}
\usepackage{MnSymbol}%
\usepackage{wasysym}%
\usepackage{cite} 
\usepackage{enumitem}
\usepackage{threeparttable} 
\usepackage[capposition=top]{floatrow}
\graphicspath{{media/}}     % organize your images and other figures under media/ folder

\title{Predicting Individualized Effects of Internet-Based Treatment for Genito-Pelvic Pain/Penetration Disorder: Development and Internal Validation of a Multivariable Decision Tree Model}

\author{
  Anna-Carlotta Zarski \\
  Friedrich-Alexander-Universität \\
  Erlangen-Nürnberg \& \\ 
  Technical University of Munich \\
  \texttt{anna.carlotta.zarski@tum.de} \\
   \And
  Mathias Harrer \\
  Technical University of Munich \& \\ 
  Friedrich-Alexander-Universität \\
  Erlangen-Nürnberg \\
  \texttt{mathias.harrer@tum.de} \\
  \AND
  Paula Kuper \\
  Otto-von-Guericke University Magdeburg \& \\
  Technical University of Munich \\
  \texttt{paula.kuper@med.ovgu.de } \\
  \And
  Antonia A. Sprenger  \\
  Otto-von-Guericke University Magdeburg \& \\
  Technical University of Munich \\
  \texttt{antonia.sprenger@med.ovgu.de } \\
  \AND
  Matthias Berking \\
  Friedrich-Alexander-Universität \\
  Erlangen-Nürnberg \\
  \texttt{matthias.berking@fau.de} \\
  \And
  David Daniel Ebert \\
  Technical University of Munich \\
  \texttt{david.daniel.ebert@tum.de} \\
}

\begin{document}

\maketitle

\begin{abstract}
Genito-Pelvic Pain/Penetration-Disorder (GPPPD) is a common disorder but rarely treated in routine care. Previous research documents that GPPPD symptoms can be treated effectively using internet-based psychological interventions. However, non-response remains common for all state-of-the-art treatments and it is unclear which patient groups are expected to benefit most from an internet-based intervention. Multivariable prediction models are increasingly used to identify predictors of heterogeneous treatment effects, and to allocate treatments with the greatest expected benefits. In this study, we developed and internally validated a multivariable decision tree model that predicts effects of an internet-based treatment on a multidimensional composite score of GPPPD symptoms. Data of a randomized controlled trial comparing the internet-based intervention to a waitlist control group ($N$=200) was used to develop a decision tree model using model-based recursive partitioning. Model performance was assessed by examining the apparent and bootstrap bias-corrected performance. The final pruned decision tree consisted of one splitting variable, joint dyadic coping, based on which two response clusters emerged. No effect was found for patients with low dyadic coping ($n$=33; $d$=0.12; 95\% CI: -0.57-0.80), while large effects ($d$=1.00; 95\%CI: 0.68-1.32; $n$=167) are predicted for those with high dyadic coping at baseline. The bootstrap-bias-corrected performance of the model was $R^2$=27.74\% (RMSE=13.22). 
\end{abstract}

\vspace{8mm}

% keywords can be removed

\newpage
\section{Introduction}

Genito-Pelvic Pain/Penetration-Disorder (GPPPD) is a sexual dysfunction previously known as dyspareunia and vaginismus, that can be defined by genito-pelvic pain, vaginal intercourse difficulties, pelvic floor muscle tension, and fear of pain or vaginal penetration (American Psychiatric Association, 2013). Overlapping conditions include Vulvodynia and Provoked Vestibulodynia (Bergeron et al., 2015). GPPPD adversely affects women’s sexuality, quality of life, physical and mental health and relationships (Arnold et al., 2006; Khandker et al., 2011; Pâquet et al., 2016; Thomtén, 2014). GPPPD symptoms are experienced by 20.8\% (1\%-72\%) of premenopausal women in the general population (McCool et al., 2016) with a 12-month prevalence rate ranging from 4.9 to 10.9 depending on its associated impairment (Briken et al., 2020). Multiple biopsychosocial factors influence the etiology of GPPPD with psychological factors such as fear avoidance and pain catastrophizing playing an important role in its maintenance (Thomtén et al., 2014; Thomtén \& Linton, 2013).

Psychological interventions have been shown to be effective in reducing genito-pelvic pain and associated distress in vaginismus and dyspareunia and enabling sexual intercourse (Flanagan et al., 2015; Maseroli et al., 2018). However, there is a lack of willingness to participate in psychological interventions due to limited availability, fear of stigmatization, and feelings of shame (Bergvall \& Himelein, 2014; Bond et al., 2015; Donaldson \& Meana, 2011). Internet-based treatment approaches are considered particularly suitable for sexual dysfunctions to make evidence-based treatment 1) more easily accessible at any time from any location, 2) well scalable in the ratio of invested resources to people reached, and 3) anonymous so that stigmatization can be reduced (Ebert et al., 2018). Meta-analytic results showed medium to large effects of internet-based treatment compared to control conditions with regard to female sexual functioning (g=0.59, CI: 0.28–0.90, I2=0\%) and satisfaction (g=0.90, CI: 0.02–1.79, I2=82\%) (Zarski et al., 2022). For GPPPD in specific, we evaluated an internet-based treatment which resulted in an improvement rate of 31\% of women being able to successfully have sexual intercourse compared to 13\% in the control group and small to medium effects on other GPPPD core symptom dimensions of genito-pelvic pain and coital and noncoital fear of sexuality (d=0.40-0.74) (Zarski et al., 2021)

However, similar to other forms of treatment for various mental disorders (Kessler et al., 2017), not all participants benefit to the same extent from internet-based treatment for sexual dysfunctions. Although individuals with mental disorders who do not receive treatment are at higher risk for deterioration, about 25\% of individuals show no changes, and 5-6\% become worse (Rozental et al., 2017, 2019). Differences in treatment outcomes can be attributed to heterogeneity in personal and disorder characteristics (Barber, 2007; Delgadillo et al., 2016; DeRubeis et al., 2014). This heterogeneity may result in different needs with regard to treatment. In order to enable appropriate treatment modality allocation as well as adapt a chosen treatment to individual needs, these varying effects should be identified (Kraemer et al., 2002). To collect a risk profile at baseline could allow different baseline conditions to be addressed in treatment, with the goal of promoting treatment success. 

Few existing studies have identified promising effect modifiers of psychological treatment outcomes for GPPPD. Higher levels of pretreatment pain intensity and catastrophizing have been shown to be associated with poorer genito-pelvic pain outcomes after psychological treatment (Bergeron et al., 2008; Brotto et al., 2015, 2020; Desrochers et al., 2010). Higher relationship satisfaction, in turn, has been found to correlate with better treatment outcome regarding sexual satisfaction and distress (Hummel et al., 2018; Stephenson et al., 2013). In contrast, one study found high joint dyadic coping and high levels in evaluation of dyadic coping associated with lower odds of sexual intercourse (Zarski et al., 2017). Results on relationship duration as predictor of pain outcomes were mixed (Brotto et al., 2020; N. O. Rosen et al., 2021). Looking at sociodemographic variables, younger age (Brotto et al., 2020; Zarski et al., 2017), and lower levels of education (Zarski et al., 2017) have been associated with better treatment outcome for GPPPD. With regard to psychological variables, lower anxiety in women has been found to be associated with pain outcomes after treatment (Brotto et al., 2020; N. O. Rosen et al., 2021). Additionally, women with higher levels of childhood maltreatment receiving CBT-based couples therapy had poorer outcomes in sexual satisfaction and sexual function at posttreatment compared to women in a lidocaine condition (Charbonneau-Lefebvre et al., 2022).

Besides some high-quality moderator analysis, most studies, however, did not include (1) specific a-priori statement about the intention of testing moderators (2) evidence- or theory-based selection of moderators (3) measurement of moderators prior to randomization (4) adequate quality of measurements of moderators, and (5) specific test of interaction between moderators and treatment (Pincus et al., 2011). Moreover, moderators were almost exclusively evaluated on a variable-by-variable basis, and more complex multivariate interactions were not taken into account. Modeling multivariate interactions within a joint prognostic model could allow to better capture the heterogeneity of treatment effects across patients (Dahabreh et al., 2016; Kent et al., 2020). In this study, we therefore aim to develop and internally validate a multivariate prognostic model predicting individualized treatment effects of an Internet-based treatment for GPPPD.

\section{Methods}
\label{sec:methods}

Where applicable, this article adheres to the "transparent reporting of a multivariable prediction model for individual prognosis or diagnosis (TRIPOD)" statement (Collins et al., 2015).

\subsection{Study Design}

The data for this study came from a two-armed randomized controlled trial ($n_{\text{intervention}}$ = 100, $n_{\text{waitlist}}$ = 100) designed to evaluate the efficacy of an internet-based treatment for GPPPD in comparison to a waitlist control condition (study registration number: DRKS00010228, ethical approval by University of Erlangen-Nürnberg: no. 324\_15B). Further information on the study and the intervention can be found in the published study protocol, a case study, and the efficacy evaluation paper (Zarski, Berking, \& Ebert, 2018; Zarski, Berking, Hannig, et al., 2018; Zarski et al., 2021). 

\subsection{Study Inclusion Criteria and Process}

Participants were women 18 years and older who were unable to have sexual intercourse for the last six months or longer and who were in a heterosexual relationship. Additionally, pre-existing medical conditions causing GPPPD symptoms had to have been ruled out before enrollment. In order to participate in the study, individuals were required to have sufficient German language skills, internet access, and provide informed consent. Those with 1) current or lifetime psychosis or dissociative symptoms, 2) present substance dependency or abuse, 3) present moderate or severe depression or bipolar disorder, 4) current or lifetime posttraumatic stress disorder or traumatization caused by sexual abuse, or 5) present treatment of GPPPD were excluded from the study. After completion of the first assessment, a 1:1 randomization through an automated computer-based random integer generator (\texttt{randlist}) took place and participants were allocated to either the intervention group (IG) or the waitlist control group (WCG). The research staff was blinded to the randomization until group allocation was successfully completed.

\subsection{Intervention}

The 10-week online intervention consisted of 9 sessions in total, including an extra booster session which took place four weeks after the program ended. These sessions incorporated psychoeducation, communication exercises, non-judgmental awareness, body exposure and genital self-exploration, attention-focusing for pain-management, cognitive reconstructing, sensate focus, gradual exposure of fingers and dilators, sexual intercourse preparation exercises, and relapse prevention. Additionally, participants kept an online diary where they monitored weekly transfer tasks as well as exercises in their daily life. They also received encouraging text messages, personalized written feedback on completed sessions, and practice reminders throughout the treatment program. Specifically, if a module had failed to be completed within seven days, a reminder from an eCoach was sent out. 

\subsection{Outcome measures}

For this study, outcome data assessed in both the IG and WCG via self-report after completion of treatment/12 weeks after randomization (T2) was used. Potential moderators were assessed prior to randomization at baseline.

\subsubsection{Multidimensional composite primary outcome measure}

\begin{figure}[H]
\includegraphics[width=4cm]{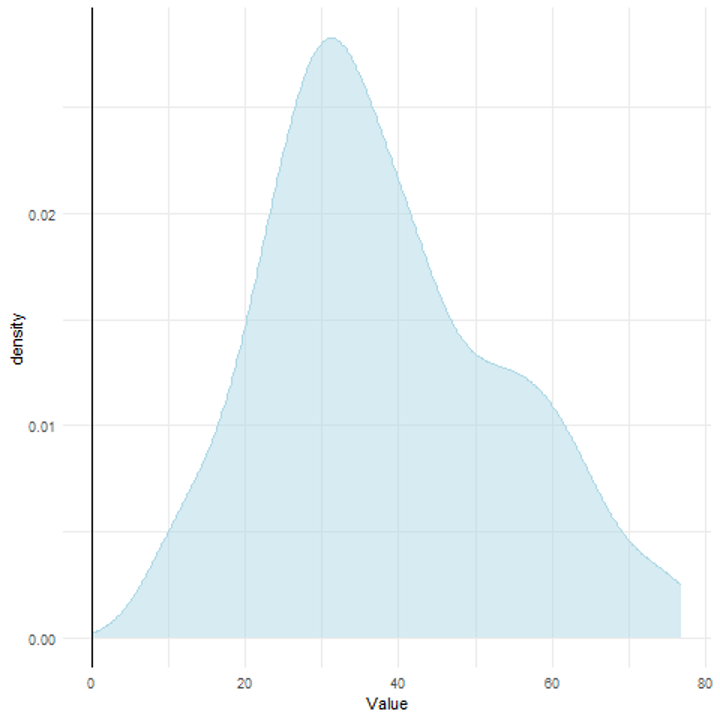}
\centering
\caption{\emph{Density plot of the aggregated outcome}}
\label{fig:fig1}
\floatfoot{\emph{Note.} Outcome at post-assessment based on imputed data.}
\end{figure}

As primary outcome, we built a composite measure to comprise the core dimensions of GPPPD by aggregating scores of the following components: coital and noncoital penetration ability, genito-pelvic pain and interference of genital pain with sexual intercourse, fear of coitus and noncoital sexual activity, and sexual satisfaction. The Primary Endpoint Questionnaire (PEQ) was used to assess intercourse penetration behavior (1 item, scores: 0 [not attempted or attempted, but unsuccessful] and 1 [attempted and sometimes successful or attempted and always successful]) and noncoital self-insertion behavior (3 items, $\alpha$=.71) (van Lankveld et al., 2006). Genito-pelvic pain (3 items, $\alpha$=.90) and sexual satisfaction (3 items, $\alpha$=.75) was measured using the respective subscales of the Female Sexual Functioning Index (FSFI) (Berner et al., 2004; R. Rosen et al., 2000). Interference of genital pain with sexual intercourse was assessed according to the Diagnostic Guidelines for the Assessment of Genito-Pelvic Pain/Penetration Disorder (3 items, $\alpha$=.60) (Binik, 2010) and fear of coitus (3 items, $\alpha$=.78) and noncoital sexual activity by the Fear of Sexuality Questionnaire (5 items, $\alpha$=.82) (FSQ) (ter Kuile et al., 2007). Comparable scaling among the included measures was achieved by 1) harmonizing measures so that higher values indicated better GPPPD symptom severity, 2) building the same number of quantiles over each scale, and 3) allocating scores in the same quantile to the same value. For the dichotomous coital penetration scale, no coital penetration was assigned to the lowest quantile and coital penetration to the highest. A solution with eleven quantiles was chosen since the primary outcome should sufficiently differentiate between participants. At the same time, the chosen number of quantiles should neither overly exceed the smallest range of the scales included nor increase the impact of the dichotomous coital penetration scale beyond measure. When computing quantiles, skewness, mean and standard deviation of the distribution of each scale were taken into account using the \textsf{R} package \texttt{sn} (Azzalini, 2021). The new scores of each scale were added to a sum score. Figure \ref{fig:fig1} shows the density plot of the aggregated outcome. As sensitivity analyses, the aggregated outcome was recalculated several times according to the leave-one-out principle, i.e., one scale score at a time was excluded to build the outcome. Respective density plots and histograms are shown in Figure \ref{fig:fig2} and \ref{fig:fig3} in the \hyperref[sec:appendix]{Appendix}.

\subsubsection{Participant-Level Moderators}

In order to examine the potential moderating role of various baseline variables relevant to GPPPD on the effect of the intervention, 39 potential moderator variables were included: nine sociodemographic variables, i.e., age (years), nationality (German/non-German), level of education (low/middle/high), married (\emph{yes/no}), children (\emph{yes/no}), previous treatment for GPPPD (\emph{yes/no}), experience with psychotherapy (\emph{yes/no}), online training experience (\emph{yes/no}); 14 variables related to sexual function, i.e., duration of GPPPD (years), lifelong GPPPD (\emph{yes/no}), sexual intercourse attempts in past 6 months, experience of sexual abuse (\emph{yes/no}), control (four items, $\alpha$=.72), catastrophic and pain (five items, $\alpha$=.73), self-image (six items, $\alpha$=.81), genital incompatibility (two items, $\alpha$=.66), and positive (five items, $\alpha$=.80) cognitions regarding vaginal penetration assessed by the Vaginal Penetration Cognition Questionnaire (VPCQ) (Klaassen \& ter Kuile, 2009), sexual desire (two items, $\alpha$=84), sexual arousal (four items, $\alpha$=93), lubrication (four items, $\alpha$=94), orgasm (three items, $\alpha$=.93) assessed by the FSFI (Berner et al., 2004; Rosen et al., 2000), noncoital insertion by the partner (three items, $\alpha$=.60) assessed by the Primary Endpoint Questionnaire (PEQ) (van Lankveld et al., 2006); 5 partnership-related variables, i.e., duration of partnership (years), delegated (two items, $\alpha$=92.), joint (five items, $\alpha$=.76), and evaluation of (two items, $\alpha$=.87) dyadic coping assessed by the Dyadic Coping Inventory (DCI) (Bodenmann, 2000; Ledermann et al., 2010), satisfaction with partnership (9 items, $\alpha$=.79) and partnership happiness (1 item) assessed by the Partnership Questionnaire Short Form (PFB-k) (Kliem et al., 2012); 4 mental health variables, i.e., self-esteem assess by the Self-Esteem Scale (ten items, SES, $\alpha$=.90) (von Collani \& Herzberg, 2003), generalized anxiety disorder symptoms assessed by the Generalized Anxiety Disorder Assessment (GAD-7, 7 items, $\alpha$=.81) (Spitzer et al., 2006), trait anxiety assessed by the State Trait Anxiety Inventory (STAI-T, 20 items, $\alpha$=.89) (Laux et al., 1981; Spielberger et al., 1970), well-being assessed by the Well-Being Index (WHO-5, $\alpha$=.84; Brähler et al., 2007); as well as the 7 GPPPD-related items used to build the aggregated outcome.

\subsection{Imputation}

Missing data at baseline and post-test was imputed under the "missing at random" assumption using random forest methodology (Stekhoven \& Bühlmann, 2012). The \textsf{R} package \texttt{missForest} (Stekhoven, 2013) is a non-parametric missing value imputation method that can handle mixed-type data on the basis of Breiman's random forests (Breiman, 2001). The imputation procedure consists of the following principal steps: First, missing values are estimated using a simple procedure (e.g., mean imputation) and the variables are sorted in ascending order according to the number of missing values. Subsequently, a random forest is fitted on the observed parts of the dataset to predict the missing values variable by variable. This is repeated iteratively until a stopping criterion is met (Stekhoven \& Bühlmann, 2012). The imputation model included all demographic and scale-level questionnaire scores collected at baseline. Additionally, variables at post-test used to build the primary outcome were included in the model. The number of trees in each forest was set to 100.

\subsection{Decision Tree Model}

Model based-recursive partitioning (MOB) (Zeileis et al., 2008) was used to examine potential treatment moderators using the \textsf{R} \texttt{partykit} package (Hothorn et al., 2015). This method can be applied when it is assumed that a single global model does not fit the data well. In such a case, MOB involved the treatment effect to be tested for heterogeneity with respect to moderator variables (via parameter stability tests). This results in a tree where each node is associated with a local model consisting of subgroups of patients with similar model trends (Zeileis et al., 2008). 

\subsection{Data Analysis}

\subsubsection{Preselection of Moderator Candidates}

Before starting MOB, potential moderator candidates are preselected in order to reduce the number of variables in the final analysis model resulting in a more parsimonious model. This was done using model-based random forest analysis (Garge et al., 2013). It allows to calculate variable permutation importance for the moderator variables through random forest methodology by constructing multiple ($n$=300) model-based trees (Garge et al., 2013). For each tree, a random subset of the splitting variables is sampled to construct the tree, resulting in more stable and less sample-specific predictions. All baseline variables except for the outcome and predictor variables are selected as potential moderator variables. Potential moderator variables are ranked in order of their contribution to the production of accurate predictions (Garge et al., 2013). 

\subsubsection{Model-Based Recursive Partitioning}

For the subsequent model-based tree analysis, variables with a positive variable importance are included as partitioning variables. The model was a linear model in which the aggregated outcome at post-treatment was regressed on the treatment indicator variable and the aggregated GPPPD symptom severity outcome at baseline. The algorithm consists of several steps that are performed iteratively over the model until no significant parameter instability with respect to the moderators can be determined anymore (Zeileis et al., 2008): 1) it fits the model to the observations in the current node, then 2) it assesses parameter instability of the treatment effect with respect to the covariates, 3) split points are computed that locally optimize the objective function and then 4) splits the current node into child nodes. By separating variable and cut-point selection, an unbiased variable selection is enabled unlike in other tree-based methods (Fokkema et al., 2021). To avoid overfitting, the significance level for parameter stability tests was set to $\alpha$=0.05 and $P$ values were Bonferroni-corrected. The minimum number of observations in each terminal node was set to 10, i.e., the default of ten times the number of estimated parameters divided by the number of responses per observation (Zeileis et al., 2008). Effect sizes were calculated in each node as Cohen's $d$ ($d$=0.2 small, $d$=0.5 medium, and $d$=0.8 large effects) with 95\% confidence intervals (CIs).

\subsubsection{Model Performance}

Model performance was assessed using (adjusted) $R^2$ and root mean squared errors (RMSE) to capture the (adjusted) proportion of variance explained by the decision tree model and the mean difference between predicted and observed values. In line with recommendations by Moons et al. (2019) and Steyerberg (2019), bootstrap bias correction was applied to internally validate the decision tree model. Via bootstrap bias correction, the proportion of excess performance that is due to overfitting and does not reflect the true population-based performance is quantified.

\section{Results}

\subsection{Participant Characteristics}

The women participating in this study were on average 28.75 (SD=8.89) years old and over half had a university degree (53.50\%, n=107). All women were in a partnership and were not able to have sexual intercourse with vaginal penetration due to GPPPD symptoms for at least the last 6 months prior to study participants (see Table \ref{tab:tab1} and Table \ref{tab:tab2}) (Zarski et al., 2021). At post-treatment, 78\% of participants in the IG and 92\% in the WCG completed the assessment. 

\begin{table}[!htbp]
\footnotesize
\caption{Participant characteristics and potential moderators at baseline.} \label{tab:tab1}
    \centering
    \begin{tabular}{llll} \toprule
    \textbf{Potential moderators}  &    \textbf{IG ($n=100$)}        & \textbf{CG ($n=100$)}   & \textbf{Total ($N=200$)}  \\\midrule
    $~$ & & & \\
    \textbf{\emph{Sociodemographic Variables}} & & & \\
    Age (years), $M$ (SD), range	         & 29.46 (9.82)	     & 28.04 (7.84)	     & 28.75 (8.89) \\
    German nationality, $n$ (\%)	         & 89 (89.00)	     & 93 (93.00)	     & 182 (91.00) \\
    Level of education, $n$ (\%)			 & & & \\
    \hspace*{5mm}- Low	                     & 0	             & 4 (4.00)	         & 4 (2.00) \\
    \hspace*{5mm}- Middle	                 & 42 (42.00)	     & 47 (47.00)	     & 89 (44.50) \\
    \hspace*{5mm}- High	                     & 58 (58.00)	     & 49 (49.00)	     & 107 (53.50) \\
    Married, $n$ (\%)	                     & 27 (27.00)	     & 24 (24.00)	     & 51 (25.50) \\
    Have children, $n$ (\%)	                 & 13 (13.00)	     & 3 (3.00)	         & 16 (8.00) \\
    Previous GPPPD treatment, $n$ (\%)	     & 32 (32.00)	     & 32 (32.00)	     & 64 (32.00) \\
    Psychotherapy experience, $n$ (\%)	     & 36 (36.00)	     & 36 (36.00)	     & 72 (36.00) \\
    Online training experience, $n$ (\%)	 & 10 (10.00)	     & 4 (4.00)	         & 14 (7.00) \\
    $~$ & & & \\ \midrule
    $~$ & & & \\
    \textbf{\emph{Sexual Function Variables}} & & & \\
    Duration of GPPPD (years), $M$ (SD)$^{\text{a}}$        & 7.90 (7.15)       & 8.13 (7.00)       & 8.02 (7.06) \\
    Lifelong GPPPD, $n$ (\%)	                            & 32 (32.00)	    & 44 (44.00)        & 76 (38.00) \\
    Intercourse attempts past 6 months	                    & 6.55 (11.29)	    & 7.34 (20.87)	    & 6.95 (16.74) \\
    Sexual abuse, $n$ (\%)$^{\text{b}}$	                    & 10 (10.99)	    & 9 (9.28)	        & 19 (10.11) \\
    Control cognitions, $M$ (SD)	                        & 2.62 (1.43)	    & 2.84 (1.46)	    & 2.73 (1.45) \\
    Catastrophic and pain cognitions, $M$ (SD)	            & 4.16 (1.18)	    & 4.09 (1.30)	    & 4.12 (1.24) \\
    Self-image cognitions, $M$ (SD)	                        & 3.28 (1.37)	    & 3.47 (1.38)	    & 3.38 (1.37) \\
    Genital incompatibility cognitions, $M$ (SD)	        & 3.05 (1.78)	    & 3.27 (1.77)	    & 3.16 (1.77) \\
    Positive cognitions, $M$ (SD)	                        & 2.26 (1.11)	    & 2.23 (1.22)	    & 2.25 (1.16) \\
    Sexual desire, $M$ (SD)	                                & 3.26 (1.06)	    & 3.35 (1.17)	    & 3.31 (1.11) \\
    Arousal, $M$ (SD)	                                    & 4.03 (1.49)	    & 3.89 (1.55)	    & 3.96 (1.52) \\
    Lubrication, $M$ (SD)	                                & 3.99 (1.63)	    & 3.91 (1.73)	    & 3.95 (1.68) \\
    Orgasm, $M$ (SD)	                                    & 3.96 (1.81)	    & 3.65 (1.82)	    & 3.81 (1.82) \\
    Noncoital insertion by the partner, $M$ (SD)	        & 0.52 (0.64)	    & 0.49 (0.63)	    & 0.50 (0.64) \\
    $~$ & & & \\ \midrule
    $~$ & & & \\
    \textbf{\emph{Partnership-Related Variables}} & & & \\
    Duration of partnership (years), $M$ (SD)	               & 6.72 (7.09)	   & 5.62 (4.59)	  & 6.17 (5.98) \\
    Delegated dyadic coping, $M$ (SD)	                       & 7.14 (2.02)	   & 7.00 (1.87)	  & 7.07 (1.94) \\
    Joint dyadic coping, $M$ (SD)	                           & 17.00 (4.04)	   & 17.26 (3.64)	  & 17.13 (3.84) \\
    Evaluation of dyadic coping, $M$ (SD)	                   & 7.37 (1.94)	   & 7.56 (1.63)	  & 7.47 (1.79) \\
    Relationship quality, $M$ (SD)	                           & 21.20 (4.21)	   & 20.86 (4.40)	  & 21.03 (4.30) \\
    Happiness in the relationship, $M$ (SD)	                   & 3.73 (0.97)	   & 3.64 (1.16)	  & 3.68 (1.07) \\
    $~$ & & & \\ \midrule
    $~$ & & & \\
    \textbf{\emph{Mental Health Variables}} & & & \\
    Self-esteem, $M$ (SD) 	             & 20.99 (5.71)      & 20.55 (5.96)	     & 20.77 (5.82) \\
    Generalized anxiety, $M$ (SD)        & 6.94 (4.05)	     & 7.65 (4.05)	     & 7.30 (4.06) \\
    Trait anxiety, $M$ (SD)	             & 50.03 (13.25)	 & 51.83 (14.47)	 & 50.93 (13.87) \\
    Well-being, $M$ (SD)	             & 53.08 (17.18)	 & 47.48 (17.40)	 & 50.28 (17.47) \\
    $~$ & & & \\ \toprule
    \end{tabular}
    \begin{tablenotes}
      \small
      \item \emph{Note.} $^{\text{a}}$refers to a subsample of $n$ = 199, $^{\text{b}}$refers to a subsample of $n$ = 188, F2F: face-to-face psychotherapy
    \end{tablenotes}
\end{table}

\begin{table}[!htbp]
\footnotesize
\caption{GPPPD-related variables at baseline and post-treatment.} \label{tab:tab2}
    \centering
    \begin{tabular}{llll} \toprule
    \textbf{GPPPD-Related Outcomes }  &    \textbf{IG ($n=100$)}        & \textbf{CG ($n=100$)}  \\\midrule
    $~$ & & & \\
    Intercourse penetration behavior T1, $n$, (\%)	& 0             & 0 \\
    Intercourse penetration behavior T2, $n$, (\%)  & 31 (31.00)    & 13 (13.00) \\
    Noncoital self-insertion T1, $M$ (SD)	        & 1.11 (0.93)	& 1.11 (0.92) \\
    Noncoital self-insertion T2, $M$ (SD)	        & 1.82 (0.85)	& 1.17 (0.91) \\
    Genito-pelvic pain T1, $M$ (SD)	                & 1.61 (1.00)	& 1.64 (1.06) \\
    Genito-pelvic pain T2, $M$ (SD)	                & 2.82 (1.36)	& 1.84 (1.26) \\
    Genital pain interference T1, $M$ (SD)	        & 4.63 (0.45)	& 4.63 (0.56) \\
    Genital pain interference T2, $M$ (SD)          & 3.72 (0.97)	& 4.37 (0.72) \\
    Coital fear T1, $M$ (SD)	                    & 11.03 (2.86)	& 11.12 (3.37) \\
    Coital fear T2, $M$ (SD)                        & 8.50 (3.20)	& 9.93 (3.43) \\
    Noncoital fear T1, $M$ (SD)	                    & 11.76 (4.39)	& 11.42 (4.09) \\
    Noncoital fear T2, $M$ (SD)	                    & 10.09 (3.55)	& 11.70 (4.51) \\
    Sexual Satisfaction T1, $M$ (SD)	            & 3.35 (1.37)	& 3.65 (1.33) \\
    Sexual Satisfaction T2, $M$ (SD)                & 3.89 (1.12)	& 3.57 (1.36) \\
    $~$ & & & \\ \toprule
    \end{tabular}
    \begin{tablenotes}
      \small
      \item \emph{Note.} T1 = baseline, T2 = post-treatment, $n$ = sample size, $M$ = mean, SD = standard deviation.
    \end{tablenotes}
\end{table}

\subsection{Decision Tree Model}

The resulting regression-based tree depicts scatter plots for GPPPD symptom severity as the aggregated outcome at baseline and at post-treatment as well as box-and-whisker plots by treatment group with GPPPD symptom severity at post-treatment as outcome measure. The data is partitioned at a joint dyadic coping index of 13 which corresponds roughly to one standard deviation below the sample mean ($M$ = 17.13; SD = 3.84). 

\subsection{Preselection of Moderator Candidates}

Three potential moderators, namely noncoital insertion by the partner, self-esteem, and interference of genital pain with sexual intercourse showed a negative variable importance and were thus excluded from the pool of partitioning variables. Variable importance was highest for joint dyadic coping with a score above 1.0. Figure \ref{fig:fig4} in the \hyperref[sec:appendix]{Appendix} illustrates the variable permutation importance of potential moderator variables using model-based random forest analysis.

\subsection{Terminal Node Models}

\begin{figure}[H]
\includegraphics[width=6cm]{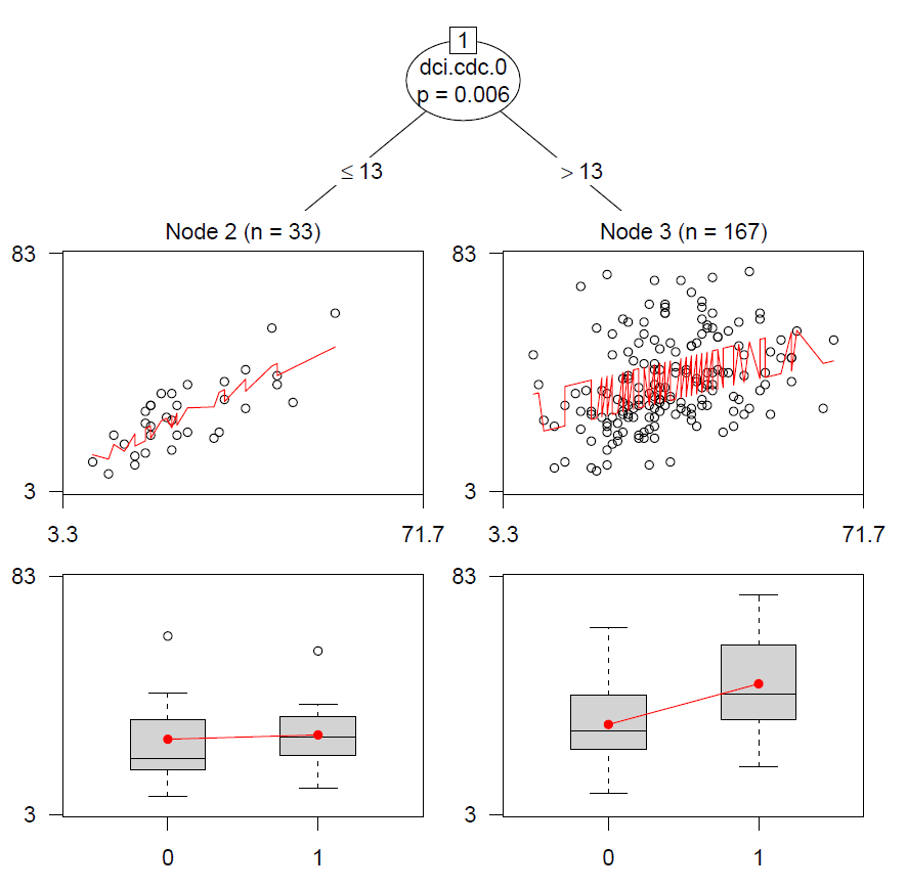}
\centering
\caption{\emph{Decision tree model; based on imputed data.}}
\label{fig:fig5}
\end{figure}

The final model-based tree is displayed in Figure \ref{fig:fig5} and the terminal node specific regression coefficients in Table 3. For women with low joint dyadic coping (Node 2; $N$ = 33), the effect of the intervention on GPPPD symptom severity was negligible ($b$ = 4.08, $p$ = 0.139) while higher GPPPD symptom severity at baseline was significantly associated with higher GPPPD symptom severity scores at post-treatment ($b$ = 0.88, $p<$0.001). In contrast, women with average-to-high joint dyadic coping (Node 3), a group comprising more than three-quarters of the participants ($N$ = 167), profited significantly from the intervention ($b$ = 13.24, $p<$0.001). Again, higher aggregated outcome scores in GPPPD symptom severity at baseline were significantly associated with higher aggregated outcome scores at post-treatment ($b$ = 0.43, $p <$0.001). 

\begin{figure}[H]
\includegraphics[width=15cm]{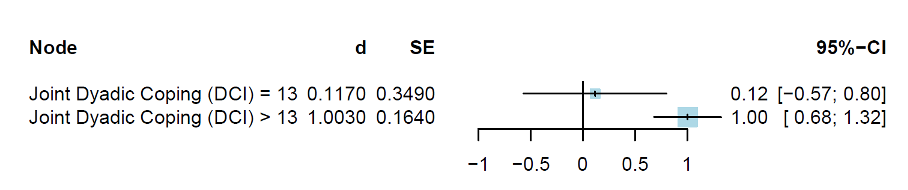}
\centering
\caption{\emph{Differential treatment effects in each terminal node resulting from the model-based decision tree}}
\floatfoot{\emph{Note.} \textsf{d} denotes Cohen's $d$ in each terminal node, \textsf{SE} denotes respective standard errors.}
\label{fig:fig6}
\end{figure}

Accordingly, standardized mean treatment effects were highest for women with average-to-high joint dyadic coping (Cohen's $d$ = 1.00, 95\%-CI: 0.68 - 1.32) with a proportion of explained variance of $R^2$ = 30.36\% (adjusted $R^2$ = 29.51). In women with low joint dyadic coping, the standardized mean treatment effect size was Cohen's $d$ = 0.12, 95\%-CI: -0.57 to 0.80. A forest plot of the subgroup-conditional effects is presented in Figure \ref{fig:fig6}.

Taken together, the model suggests that women with average-to-high joint dyadic coping will significantly benefit from the intervention in respect to their GPPPD symptom severity while women with low joint dyadic coping will probably not benefit. Assessing joint dyadic coping at an early stage might thus be helpful for beneficial treatment assignment.

\subsection{Model Performance}

The proportion of variance explained by the decision tree model was $R^2$ = 38.64\% (adjusted $R^2$ = 38.33\%). After bootstrap bias correction, the proportion of explained variance was reduced by about 10\% ($R^2$ = 27.74\%; adjusted $R^2$ = 27.38\%). The mean difference between predicted and observed values was RMSE = 11.93. After bootstrap bias correction, this difference was slightly increased (RMSE = 13.22). Both measures indicated substantial predictive performance of the decision tree model even after bootstrap bias correction.

\subsection{Planned Further Analyses}

To draw firm conclusions about this prediction model of individualized effects of Internet-based treatment for GPPPD and to derive evidence-based practical implications, external validation of the multivariable decision tree model is needed. Therefore, the second aim of this study is to validate the tree-based prediction model in a second, independent randomized-controlled trial on the evaluation of internet-based treatment for GPPPD.

\begin{flushright}
    $\blacksquare$
\end{flushright}

\section*{Funding}

The project was supported by the "Gender \& Diversity" funding of the Friedrich-Alexander-Universität Erlangen-Nürnberg.

\section*{Supplementary Information}

Data is available upon reasonable request from the first author.  

\newpage

\section*{References}

\begin{itemize}[label={},itemindent=-2em,leftmargin=2em]

\item American Psychiatric Association. (2013). Diagnostic and Statistical Manual of Mental Disorders (5th ed.). American Psychiatric Association.

\item Arnold, L. D., Bachmann, G. A., Rosen, R., Kelly, S., \& Rhoads, G. G. (2006). Vulvodynia: characteristics and associations with comorbidities and quality of life. \emph{Obstetrics and Gynecology, 107}(3), 617–624. https://doi.org/10.1097/01.AOG.0000199951.26822.27 

\item Azzalini, A. (2021). \emph{sn: The Skew-Normal and Related Distributions Such as the Skew-t and the SUN}. R package version 2.0.0.

\item Barber, J. P. (2007). Issues and findings in investigating predictors of psychotherapy outcome: Introduction to the special section. In \emph{Psychotherapy Research} (Vol. 17, Issue 2, pp. 131–136). https://doi.org/10.1080/10503300601175545  

\item Bergeron, S., Corsini-Munt, S., Aerts, L., Rancourt, K., \& Rosen, N. O. (2015). Female sexual pain disorders: A review of the literature on etiology and treatment. \emph{Current Sexual Health Reports, 1–11}. https://doi.org/10.1007/s11930-015-0053-y  

\item Bergeron, S., Khalifé, S., Glazer, H. I., \& Binik, Y. M. (2008). Surgical and behavioral treatments for vestibulodynia. Two-and-one-half year follow-up and predictors of outcome. \emph{Obstetrics and Gynecology, 111}(1), 159–166. https://doi.org/10.1097/01 .AOG.0000295864.76032.a7 

\item Bergvall, L., \& Himelein, M. J. (2014). Attitudes toward seeking help for sexual dysfunctions among US and Swedish college students. \emph{Sexual and Relationship Therapy, 29}(2), 215–228. https://doi.org/10.1080/14681994.2013.860222  

\item Berner, M. M., Kriston, L., Zahradnik, H.-P., Härter, M., \& Rohde, A. (2004). Überprüfung der Gültigkeit und Zuverlässigkeit des Deutschen Female Sexual Function Index (FSFI-d). \emph{Geburtshilfe Und Frauenheilkunde, 64}(3), 293–303. https://doi.org/10.1055/s-2004-815815  

\item Binik, Y. M. (2010). The DSM diagnostic criteria for vaginismus. \emph{Archives of Sexual Behavior, 39}(2), 278–291. https://doi.org/10.1007/s10508-009-9560-0  

\item Bodenmann, G. (2000). \emph{Stress und Coping bei Paaren} [Stress and coping in couples]. Hogrefe.

\item Bond, K., Mpofu, E., \& Millington, M. (2015). Treating women with genito-pelvic pain/penetration disorder: Influences of patient agendas on help-seeking. \emph{Journal of Family Medicine, 2}(4), 1–8. issn: 2380-0658

\item Brähler, E., Mühlan, H., Albani, C., \& Schmidt, S. (2007). Teststatistische Prüfung und Normierung der Deutschen Versionen des EUROHIS-QOL Lebensqualität-index und des WHO-5 Wohlbefindens-index. \emph{Diagnostica, 53}(2), 83–96. https://doi.org/10.1026/0012-1924.53.2.83   
Breiman, L. (2001). Random Forests (Vol. 45).

\item Briken, P., Matthiesen, S., Pietras, L., Wiessner, C., Klein, V., Reed, G. M., \& Dekker, A. (2020). Prävalenzschätzungen sexueller Dysfunktionen anhand der neuen ICD-11-Leitlinien. \emph{Deutsches Ärzteblatt International, 117}(39), 653–658. https://doi.org/10.3238/arztebl.2020.0653 

\item Brotto, L. A., Basson, R., Smith, K. B., Driscoll, M., \& Sadownik, L. (2015). Mindfulness-based Group Therapy for Women with Provoked Vestibulodynia. \emph{Mindfulness, 6}(3), 417–432. https://doi.org/10.1007/s12671-013-0273-z  

\item Brotto, L. A., Zdaniuk, B., Rietchel, L., Basson, R., \& Bergeron, S. (2020). Moderators of Improvement from Mindfulness-Based vs Traditional Cognitive Behavioral Therapy for the Treatment of Provoked Vestibulodynia. \emph{Journal of Sexual Medicine, 17}(11), 2247–2259. https://doi.org/10.1016/j .jsxm.2020.07.080 

\item Charbonneau-Lefebvre, V., Vaillancourt-Morel, M. P., Rosen, N. O., Steben, M., \& Bergeron, S. (2022). Attachment and Childhood Maltreatment as Moderators of Treatment Outcome in a Randomized Clinical Trial for Provoked Vestibulodynia. \emph{Journal of Sexual Medicine, 19}(3), 479–495. https://doi.org/10.1016/j.jsxm.2021.12.013  

\item Collins, G. S., Reitsma, J. B., Altman, D. G., \& Moons, K. G. M. (2015). Transparent reporting of a multivariable prediction model for individual prognosis or diagnosis (TRIPOD): The TRIPOD Statement. \emph{BMC Medicine, 13}(1). https://doi.org/10.1186/s12916-014-0241-z   

\item Dahabreh, I. J., Hayward, R., \& Kent, D. M. (2016). Using group data to treat individuals: Understanding heterogeneous treatment effects in the age of precision medicine and patient-centred evidence. \emph{International Journal of Epidemiology, 45}(6), 2184–2193. https://doi.org/10.1093/ije/dyw125  

\item Delgadillo, J., Moreea, O., \& Lutz, W. (2016). Different people respond differently to therapy: A demonstration using patient profiling and risk stratification. \emph{Behaviour Research and Therapy, 79}, 15–22. https://doi.org/10.1016/j.brat.2016.02.003   

\item DeRubeis, R. J., Gelfand, L. A., German, R. E., Fournier, J. C., \& Forand, N. R. (2014). Understanding processes of change: How some patients reveal more than others-and some groups of therapists less-about what matters in psychotherapy. \emph{Psychotherapy Research, 24}(3), 419–428. https://doi.org/10.1080/10503307.2013.838654    

\item Desrochers, G., Bergeron, S., Khalifé, S., Dupuis, M. J., \& Jodoin, M. (2010). Provoked vestibulodynia: Psychological predictors of topical and cognitive-behavioral treatment outcome. \emph{Behaviour Research and Therapy, 48}(2), 106–115. https://doi.org/10.1016/j.brat.2009.09.014    

\item Donaldson, R. L., \& Meana, M. (2011). Early dyspareunia experience in young women: Confusion, consequences, and help-seeking barriers. \emph{Journal of Sexual Medicine, 8}(3), 814–823. https://doi.org/10.1111/j.1743-6109.2010.02150.x   

\item Ebert, D. D., van Daele, T., Nordgreen, T., Karekla, M., Compare, A., Zarbo, C., Brugnera, A., Øverland, S., Trebbi, G., Jensen, K. L., Kaehlke, F., Baumeister, H., \& Taylor, J. (2018). Internet and mobile-based psychological interventions: Applications, efficacy and potential for improving mental health. A report of the EFPA E-Health Taskforce. \emph{European Psychologist, 23}(2), 167–187. https://doi.org/10.1027/1016-9040/a000346   

\item Flanagan, E., Herron, K. A., O’Driscoll, C. C., \& de Williams, A. C. C. C. (2015). Psychological treatment for vaginal pain: Does etiology matter? A systematic review and meta-analysis. \emph{Journal of Sexual Medicine, 12}(1), 3–16. https://doi.org/10.1111/jsm.12717  

\item Fokkema, M., Edbrooke-Childs, J., \& Wolpert, M. (2021). Generalized linear mixed-model (GLMM) trees: A flexible decision-tree method for multilevel and longitudinal data. \emph{Psychotherapy Research, 31}(3), 329–341. https://doi.org/10.1080/10503307.2020.1785037   

\item Garge, N. R., Bobashev, G., \& Eggleston, B. (2013). \emph{Random forest methodology for model-based recursive partitioning: the mobForest package for R}.

\item Hothorn, T., Zeileis, A., Cheng, E., \& Ong, S. (2015). partykit: A Modular Toolkit for Recursive Partytioning in R. In \emph{Journal of Machine Learning Research} (Vol. 16). 

\item Hummel, S. B., van Lankveld, J. J., Oldenburg, H. S. A., Hahn, D. E. E., Broomans, E., \& Aaronson, N. K. (2018). Internet-based Cognitive Behavioral Therapy for DSM-IV Sexual Dysfunctions in Breast Cancer Survivors: Predictors of Treatment Response. \emph{International Journal of Sexual Health, 30}(3), 281–294. https://doi.org/10.1080/19317611.2018.1491925    

\item Kent, D. M., van Klaveren, D., Paulus, J. K., D’Agostino, R., Goodman, S., Hayward, R., Ioannidis, J. P. A., Patrick-Lake, B., Morton, S., Pencina, M., Raman, G., Ross, J. S., Selker, H. P., Varadhan, R., Vickers, A., Wong, J. B., \& Steyerberg, E. W. (2020). The Predictive Approaches to Treatment effect Heterogeneity (PATH) statement: Explanation and elaboration. \emph{Annals of Internal Medicine, 172}(1), W1–W25. https://doi.org/10.7326/M18-3668   

\item Kessler, R. C., van Loo, H. M., Wardenaar, K. J., Bossarte, R. M., Brenner, L. A., Ebert, D. D., de Jonge, P., Nierenberg, A. A., Rosellini, A. J., Sampson, N. A., Schoevers, R. A., Wilcox, M. A., \& Zaslavsky, A. M. (2017). Using patient self-reports to study heterogeneity of treatment effects in major depressive disorder. \emph{Epidemiology and Psychiatric Sciences, 26}(1), 22–36. https://doi.org/10.1017/S2045796016000020   

\item Khandker, M., Brady, S. S., Vitonis, A. F., MacLehose, R. F., Stewart, E. G., \& Harlow, B. L. (2011). The Influence of depression and anxiety on risk of adult onset vulvodynia. \emph{Journal of Women’s Health, 20}(10), 1445–1451. https://doi.org/10.1089/jwh.2010.2661  

\item Klaassen, M., \& ter Kuile, M. M. (2009). Development and initital validation of the vaginal penetration cognition questionnaire (VPCQ) in a sample of women with vaginismus and dyspareunia. \emph{J Sex Med, 6}(6), 1617–1627. https://doi.org/10.1111/j.1743-6109.2009.01217.x  

\item Kliem, S., Job, A.-K., Kröger, C., Bodenmann, G., Stöbel-Richter, Y., Hahlweg, K., \& Brähler, E. (2012). Entwicklung und Normierung einer Kurzform des Partnerschaftsfragebogens (PFB-K) an einer repräsentativen deutschen Stichprobe. \emph{Zeitschrift Für Klinische Psychologie Und Psychotherapie, 41}(2), 81–89. https://doi.org/10.1026/1616-3443/a000135    

\item Laux, L., Glanzmann, P., Schaffner, P., \& Spielberger, C. D. (1981). Das State-Trait-Angstinventar (Testmappe mit Handanweisung, Fragebogen STAI-G Form X 1 und Fragebogen STAI-G Form X 2). Beltz.

\item Ledermann, T., Bodenmann, G., Gagliardi, S., Charvoz, L., Verardi, S., Rossier, J., Bertoni, A., \& Iafrate, R. (2010). Psychometrics of the dyadic coping inventory in three language groups. \emph{Swiss Journal of Psychology, 69}(4), 201–212. https://doi.org/10.1024/1421-0185/a000024    

\item Maseroli, E., Scavello, I., Campone, B., Di Stasi, V., Cipriani, S., Felciai, F., Camartini, V., Magini, A., Castellini, G., Ricca, V., Maggi, M., \& Vignozzi, L. (2018). Psychosexual Correlates of Unwanted Sexual Experiences in Women Consulting for Female Sexual Dysfunction According to Their Timing Across the Life Span. \emph{Journal of Sexual Medicine, 15}(12), 1739–1751. https://doi.org/10.1016/j   .jsxm.2018.10.004 

\item McCool, M. E., Zuelke, A., Theurich, M. A., Knuettel, H., Ricci, C., \& Apfelbacher, C. (2016). Prevalence of female sexual dysfunction among premenopausal women: A systematic review and meta-analysis of observational studies. \emph{Sexual Medicine Reviews, 4}(3), 197–212. https://doi.org/10.1016/j.sxmr.2016.03.002  

\item Moons, K. G. M., Wolff, R. F., Riley, R. D., Whiting, P. F., Westwood, M., Collins, G. S., Reitsma, J. B., Kleijnen, J., \& Mallett, S. (2019). PROBAST: A tool to assess risk of bias and applicability of prediction model studies: Explanation and elaboration. In \emph{Annals of Internal Medicine} (Vol. 170, Issue 1, pp. W1–W33). American College of Physicians. https://doi.org/10.7326/M18-1377   

\item Pâquet, M., Bois, K., Rosen, N. O., Mayrand, M.-H., Charbonneau-Lefebvre, V., \& Bergeron, S. (2016). Why us? Perceived injustice is associated with more sexual and psychological distress in couples coping with genito-pelvic pain. \emph{The Journal of Sexual Medicine, 13}(1), 79–87. https://doi.org/10.1016/j.jsxm.2015.11.007  

\item Pincus, T., Miles, C., Froud, R., Underwood, M., Carnes, D., \& Taylor, S. J. (2011). Methodological criteria for the assessment of moderators in systematic reviews of randomised controlled trials: A consensus study. \emph{BMC Medical Research Methodology, 11}. https://doi.org/10.1186/1471-2288-11-14   

\item Rosen, N. O., Vaillancourt-Morel, M. P., Corsini-Munt, S., Steben, M., Delisle, I., Baxter, M. lou, \& Bergeron, S. (2021). Predictors and Moderators of Provoked Vestibulodynia Treatment Outcome Following a Randomized Trial Comparing Cognitive-Behavioral Couple Therapy to Overnight Lidocaine. \emph{Behavior Therapy}. https://doi.org/10.1016/j.beth.2021.05.002  

\item Rosen, R., Brown, C., Heiman, J., Leiblum, S., Meston, C., Shabsigh, R., Ferguson, D., \& D’Agostino, R. (2000). The female sexual function index (FSFI): A multidimensional self-report instrument for the assessment of female sexual function. \emph{Journal of Sex \& Marital Therapy, 26}(2), 191–208. https://doi.org/10.1080/009262300278597   

\item Rozental, A., Andersson, G., \& Carlbring, P. (2019). In the absence of effects: An individual patient data meta-analysis of non-response and its predictors in internet-based cognitive behavior therapy. \emph{Frontiers in Psychology}, 10.

\item Rozental, A., Magnusson, K., Boettcher, J., Andersson, G., \& Carlbring, P. (2017). For Better or Worse: An Individual Patient Data Meta-Analysis of Deterioration Among Participants Receiving Internet-Based Cognitive Behavior Therapy. \emph{Journal of Consulting and Clinical Psychology}, 85(2), 160–177. https://doi.org/10.1037/ccp0000158   

\item Spielberger, C. D., Gorsuch, R. L., \& Lushene, R. E. (1970). \emph{State-Trait Anxiety Inventory, Manual for the State-Trait Anxiety Inventory}. Consulting Psychologist Press.

\item Spitzer, R. L., Kroenke, K., Williams, J. B., \& Löwe, B. (2006). A brief measure for assessing generalized anxiety disorder. \emph{Arch Intern Med, 166}, 1092–1097. https://doi.org/10.1001/archinte.166.10.1092 

\item Stekhoven, D. J. (2013). \emph{missForest: Nonparametric Missing Value Imputation using Random Forest}. R package version 1.4.

\item Stekhoven, D. J., \& Bühlmann, P. (2012). Missforest-Non-parametric missing value imputation for mixed-type data. \emph{Bioinformatics, 28(1)}, 112–118. https://doi.org/10.1093/bioinformatics/btr597 

\item Stephenson, K. R., Rellini, A. H., \& Meston, C. M. (2013). Relationship satisfaction as a predictor of treatment response during cognitive behavioral sex therapy. \emph{Archives of Sexual Behavior, 42}(1), 143–152. https://doi.org/10.1007/s10508-012-9961-3   

\item Steyerberg, E. W. (2019). Clinical Prediction Models: A Practical Approach to Development, Validation, and Updating (2nd ed.). http://www.springer.com/series/2848

\item ter Kuile, M. M., van Lankveld, J. J., Groot, E. d., Melles, R., Neffs, J., Zandbergen, M., de Groot, E., Melles, R., Neffs, J., \& Zandbergen, M. (2007). Cognitive-behavioral therapy for women with lifelong vaginismus: Process and prognostic factors. \emph{Behaviour Research and Therapy, 45}(2), 359–373. https://doi.org/10.1016/j.brat.2006.03.013  

\item Thomtén, J. (2014). Living with genital pain: Sexual function, satisfaction, and help-seeking among women living in Sweden. \emph{Scandinavian Journal of Pain, 5}(1), 19–25. https://doi.org/10.1016/j.sjpain.2013.10.002  

\item Thomtén, J., \& Linton, S. J. (2013). A psychological view of sexual pain among women: Applying the fear-avoidance model. \emph{Women’s Health, 9}(3), 251–263. https://doi.org/10.2217/whe.13.19  

\item Thomtén, J., Lundahl, R., Stigenberg, K., \& Linton, S. (2014). Fear avoidance and pain catastrophizing among women with sexual pain. \emph{Women’s Health, 10}(6), 571–581.

\item van Lankveld, J. J., Melles, R., Zandbergen, M., ter Kuile, M. M., de Groot, H. E., \& Nefs, J. (2006). Cognitive-behavioral therapy for women with lifelong vaginismus: A randomized waiting-list controlled trial of efficacy. \emph{Journal of Consulting and Clinical Psychology, 74}(1), 168–178. https://doi.org/10.1037/0022-006X.74.1.168   

\item von Collani, G., \& Herzberg, P. Y. (2003). Eine revidierte Fassung der deutschsprachigen Skala zum Selbstwertgefühl von Rosenberg. Zeitschrift Für Differentielle Und Diagnostische Psychologie, 24(1), 3–7. https://doi.org/10.1024//0170-1789.24.1.3   

\item Zarski, A.-C., Berking, M., \& Ebert, D. D. (2018). Efficacy of internet-based guided treatment for genito-pelvic pain/penetration disorder: Rationale, treatment protocol, and design of a randomized controlled trial. \emph{Frontiers in Psychiatry, 22}(8), 260. https://doi.org/10.3389/fpsyt.2017.00260  

\item Zarski, A.-C., Berking, M., \& Ebert, D. D. (2021). Efficacy of internet-based treatment for genito-pelvic pain/penetration disorder: Results of a randomized controlled trial. \emph{Journal of Consulting and Clinical Psychology, 89}(11), 909–924. https://doi.org/10.1037/ccp0000665  

\item Zarski, A.-C., Berking, M., Fackiner, C., Rosenau, C., \& Ebert, D. D. (2017). Internet-based guided self-help for vaginal penetration difficulties: Results of a randomized controlled pilot trial. \emph{The Journal of Sexual Medicine, 14}(2), 238–254. https://doi.org/10.1016/j  .jsxm.2016.12.232 

\item Zarski, A.-C., Berking, M., Hannig, W., \& Ebert, D. D. (2018). Internet-Based Treatment for Genito-Pelvic Pain/Penetration Disorder: A Case Report. \emph{Verhaltenstherapie, 28}(3), 177–184. https://doi.org/10.1159/000485041    

\item Zarski, A.-C., Velten, J., Knauer, J., Berking, M., \& Ebert, D. D. (2022). Internet- and mobile-based psychological interventions for sexual dysfunctions: a systematic review and meta-analysis. In \emph{npj Digital Medicine} (Vol. 5, Issue 1). Nature Research. https://doi.org/10.1038/s41746-022-00670-1    

\item Zeileis, A., Hothorn, T., \& Hornik, K. (2008). Model-based recursive partitioning. \emph{Journal of Computational and Graphical Statistics}, 17(2), 492–514. https://doi.org/10.1198/106186008X319331      
 
\end{itemize}

\newpage

\section*{Appendix}
\label{sec:appendix}

\begin{figure}[H]
\includegraphics[width=7cm]{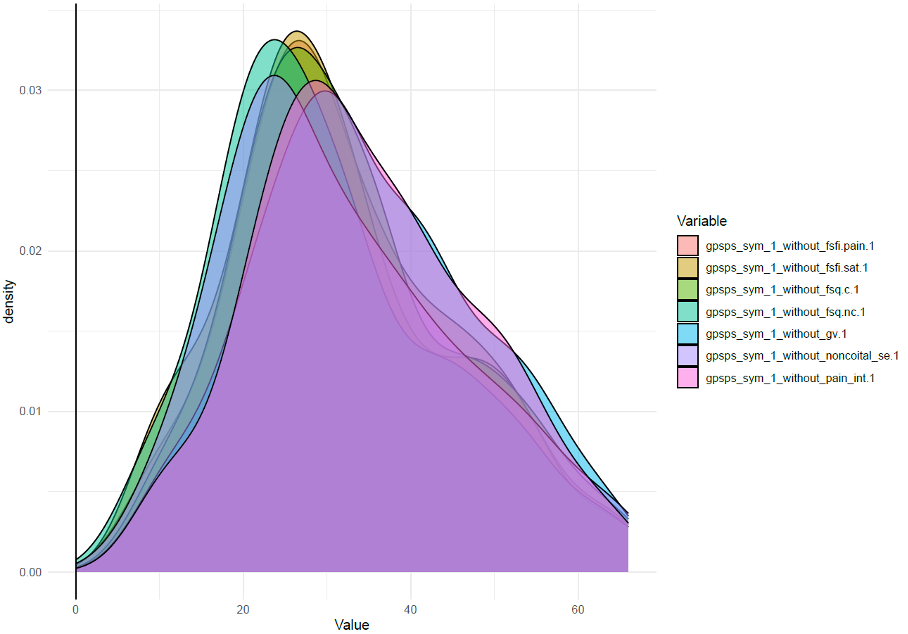}
\centering
\caption{\emph{Leave-one-out density plots of the aggregated outcome if one scale score at a time is excluded to build the outcome; based on imputed data.}}
\label{fig:fig2}
\end{figure}

\begin{figure}[H]
\includegraphics[width=8cm]{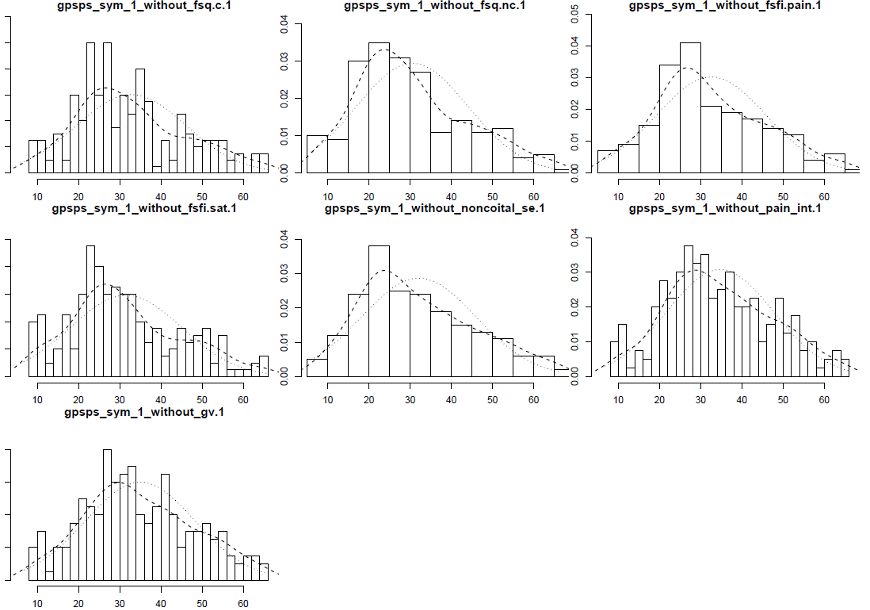}
\centering
\caption{\emph{Leave-one-out histograms of the aggregated outcome if one scale score at a time is excluded to build the outcome; based on imputed data.}}
\label{fig:fig3}
\end{figure}

\begin{figure}[H]
\includegraphics[width=6cm]{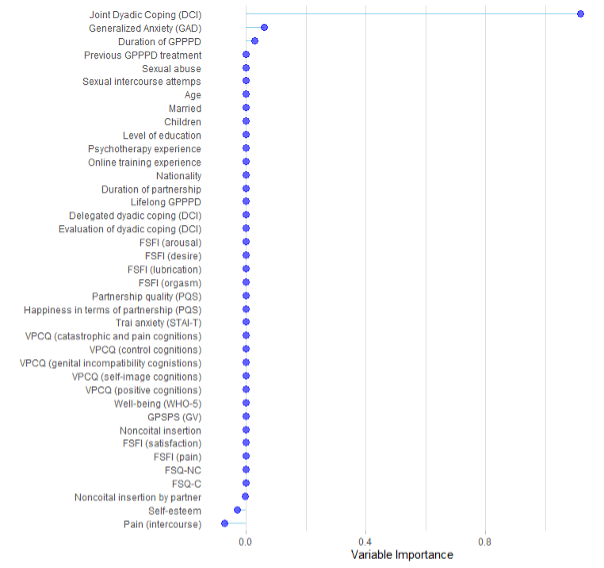}
\centering
\caption{\emph{Variable permutation importance of potential moderator variables using model-based random forest analysis; based on imputed data.}}
\label{fig:fig4}
\end{figure}

\begin{table}
    \caption{Regression coefficients of composite scores at post-treatment on composite scores at baseline and group in terminal node models}
    \centering
    \begin{tabular}{lcccc}
        \toprule
		Variable & Estimate & \emph{SE} & \emph{t} & \emph{p} \\
		\midrule
        \multicolumn{1}{l}{Node 2 $^a$} \\
		\hspace{0.5cm} Constant & 3.435 & 3.982 & 0.863 & .395 \\
        \hspace{0.5cm} Composite Score at Baseline & 0.877 & 0.122 & 7.189 & <.001 \\
        \hspace{0.5cm} Group $^c$ & 4.084 & 2.686 & 1.520 & .139 \\
        \midrule
        \multicolumn{1}{l}{Node 3 $^b$} \\
        \hspace{0.5cm} Constant & 18.694 & 3.301 & 5.663 & <.001 \\
        \hspace{0.5cm} Composite Score at Baseline & 0.428 & 0.088 & 4.873 & <.001 \\
        \hspace{0.5cm} Group $^c$ & 13.243 & 1.978 & 6.695 & <.001 \\
        \bottomrule
    \end{tabular}
    \begin{tablenotes}
        \item \textit{Note.} Node 2 comprises participants with low, Node 3 participants with average-to-high joint dyadic coping. 
        \item $^\text{a}$\textit{N} = 33. $^\text{b}$\textit{N} = 167. $^\text{c}$0 = waitlist control group, 1 = intervention group. 
    \end{tablenotes}
\end{table}

\end{document}